\documentclass[11pt,preprint,showpacs]{revtex4}
\usepackage{graphicx}
\usepackage{dcolumn}
\usepackage{bm}
\usepackage{amsmath}
\usepackage{amssymb}
\usepackage{amsfonts}

\begin{document}

\title{Dineutron and the three-nucleon continuum observables}

\author{H.~Wita{\l}a}
\affiliation{M. Smoluchowski Institute of Physics, Jagiellonian
University,
                    PL-30059 Krak\'ow, Poland}

\author{W.\ Gl\"ockle}
\affiliation{Institut f\"ur theoretische Physik II,
Ruhr-Universit\"at Bochum, D-44780 Bochum, Germany}

\date{\today}

\begin{abstract}
We investigate how strong a hypothetical $^1S_0$ bound state of two
neutrons would affect different observables in the neutron-deuteron
reactions. To that aim we extend our momentum space scheme of solving
three-nucleon Faddeev equations to incorporate in addition to the
deuteron also the $^1S_0$ dineutron bound state. We discuss effects
induced by dineutron on the angular distribution of the 
neutron-deuteron elastic scattering and cross sections of the 
deuteron breakup. 
 A comparison to the available data for neutron-deuteron total cross
 sections and  elastic scattering angular distributions 
  cannot decisively  exclude a possibility that the two neutrons 
 can form $^1S_0$ bound state. However, the strong modifications of a
 final-state-interaction peak of the neutron-deuteron breakup when
 changing from negative to positive values of the 
 neutron-neutron scattering length
 seems to exclude existence of dineutron.
\end{abstract}


\maketitle \setcounter{page}{1}

\section{Introduction}
 \label{intro}

Investigation of the neutron-deuteron (nd) elastic scattering and 
 the deuteron breakup reaction~\cite{physrep96} revealed a 
number of discrepancies between 
data and theoretical predictions based on modern nucleon-nucleon
potentials such as AV18~\cite{av18}, CD~Bonn~\cite{cdbonn}, and Nijm1,
 2 and 93~\cite{nijm}  or on the nuclear forces
derived in the framework of 
chiral perturbation theory~\cite{epel}. These potentials 
describe
very accurately all existing nucleon-nucleon (NN) data as expressed by
the value of $\chi^2$ per data point close to $\approx 1$. Some of
 these discrepancies can be explained when in adddition to pairwise
interactions also three-nucleon forces (3NF's) are included in the 3N
Hamiltonian. However, some persistently avoid explanation
and they reveal high insensitivity to the underlying dynamics, especially to
different choices among available 3NF's. The neutron-neutron (nn)
quasi-free-scattering (QFS) configuration in the complete nd breakup
is one such an example. Another is the symmetrical space-star (SST)
geometry in that reaction. The nn QFS refers to the 
kinematical configuration in which the outgoing proton is at rest in the
laboratory system. In case of SST  three outgoing nucleons have equal
magnitudes of 
momenta, which in the three-nucleon c.m. system form a plane
perpendicular to the incoming nucleon momentum with the angle of
$120^o$ between two consequitive momenta. In QFS and SST configurations
theoretical predicitons drastically underestimate  by about $\approx 20
\%$ the data. This together with the fact that cross sections in these
configurations are dominated 
by the $^1S_0$ and $^3S_1$ contributions  lead
to suspect that something is wrong with the $^1S_0$ nn force and that
maybe two neutrons can  even form a bound state~\cite{wit1,wit2}. 

That motivated us to investigate what consequences the existence of
 the $^1S_0$ dineutron would have on different observables in nd reactions
and to what degree the available data for nd reactions allow for such
 a bound state. Also we would like to see if existence of dineutron could
help in resolving the discrepancies present in QFS and SST nd breakup
configurations.

In section \ref{sec1} we extend our formulation of the momentum space
treatment of the 3N Faddeev equations to include, in addition to the
deuteron, also the $^1S_0$ bound state of two neutrons. Changing the
strength of $^1S_0$ nn interaction of the CD~Bonn potential
we produce a number of forces which allow for two neutrons being 
bound with different dineutron binding
energy. In section \ref{sec2}  we present 
 theoretical predictions based on solution of 3N Faddeev equations 
and compare them to the available nd data. We summarize and conclude 
 in section \ref{sec7}.

\section{Fadeev equations with dineutron}
\label{sec1}

We shortly present the basics of our momentum space treatment of 3N
Faddeev equations and calculations of the transition operators for
different reactions in the 3N continuum based on solutions of these
equations. For detailed presentation we refer to
\cite{physrep96,gloeckle83}. We put emphasis on changes in the
standard approach in the case when, beside the deuteron, also
one additional bound state appears in some partial wave.

For calculation of processes initiated from a state $\vert \Phi_{1,1}
> \equiv \vert \vec q_0, \phi_d >$, which  describes the neutron moving with
the relative momentum $\vec q_0$ with respect to the deuteron of the
wave function $\phi_d$, one needs the state $\vert T >$
which fulfills 3N Fadddeev equation
\begin{eqnarray}
|T> =  t P | \Phi_{1,1}> +  t P G_0 |T> ~,
\label{eq1}
\end{eqnarray} 
where $ P $ is defined in terms of transposition operators of three
nucleons, $ P =P_{12} P_{23} + P_{13} P_{23} $, $G_0$ is the free 3N propagator,
 and t is the two-nucleon off-shell t-matrix.
 Knowing $ |T>  $ the breakup as well as the elastic nd scattering 
 amplitudes
 can be gained by quadratures in the standard
 manner~\cite{physrep96}. Namely, the transition amplitude for 
 the elastic scattering, $<\Phi_{1,1}'|U|\Phi_{1,1}>$, is given
by~\cite{gloeckle83,physrep96}
\begin{eqnarray}
 \left\langle {\Phi_{1,1} '} | U | {\Phi_{1,1} } \right\rangle
 &=&
 \left\langle {\Phi _{1,1} '} | PG_0^{ - 1} | {\Phi _{1,1} } 
\right\rangle + \left\langle {\Phi_{1,1} '} | P |T \right\rangle ~, 
\label{eq2}
\end{eqnarray} 
and for the breakup, $<\Phi_0|U_0|\Phi_{1,1}>$, by
\begin{eqnarray}
 \left\langle {\Phi _0 } \right|U_0 \left| {\Phi _{1,1} } \right\rangle  &=&
 \left\langle {\Phi _0 } \right|(1 + P)\left|T \right\rangle
 ~.
\label{eq3}
\end{eqnarray}
The state $\left | {\Phi _0 } \right \rangle \equiv \frac {1}
{\sqrt{2}} (1 - P_{23})\left | \vec p \vec q~ \right \rangle$ 
corresponds to a kinematically complete configuration 
 of the breakup
described by standard Jacobi momenta $\vec p$ and $\vec q$, 
and  $ | \Phi_{1,1}'>$ is the
outgoing state of the elastic scattering with changed direction of the 
relative neutron-deuteron momentum $\vec q_0~'$ but with the same magnitude
 as in the entrance channel  $| \vec q_0~' |= | \vec q_0 |$.

Introducing the   momentum space 3N partial wave basis
     $ | pq \alpha> \equiv | pq (ls)j (\lambda 1/2) I (jI) J (t1/2)T>$ 
  with the two-body subsystem angular momenta, spin and isospin $(ls)j$
  and $t$, coupled together with the corresponding quantum numbers of
  the spectator nucleon $(\lambda 1/2)I$ and $1/2$ to the total
  angular momentum $J$ and isospin $T$ of the 3N system,   
 and projecting Eq.~(\ref{eq1}) on these states, we
    get the system of coupled integral equations in two continuous
    variables p and q. For details of the numerical treatment of that
    system, and particularly of the 
    kernel part $\langle p q \alpha \vert  t P G_0 \vert T \rangle$,   
we refer to~\cite{physrep96}. 

The 2-nucleon t-matrix conserves the spectator momentum
 $q $ and all discrete quantum numbers except the orbital angular
 momentum $l$:
\begin{eqnarray}
  < pq \alpha| t | p~ ' q~ ' \alpha~ '> &=& \frac{\delta( q - q
 ')}{q^2}t_{l_{\alpha}l_{\bar \alpha}}^{s_{\alpha} j_{\alpha}
 t_{\alpha}}( p p~ '; E(q) = E-\frac{3}{4m} q^2)  \cr
 &&  \delta_{s_{\alpha} s_{\alpha~ '}} \delta_{j_{\alpha} j_{\alpha~'}}
\delta_{t_{\alpha}
t_{\alpha~'}}\delta_{\lambda_{\alpha}\lambda_{\alpha~'}}
\delta_{I_{\alpha} I_{\alpha~'}} 
\label{eq4}
\end{eqnarray}
and has a pole in channels $\alpha$ for which two-nucleon subsystem
has bound state. 

In the channels $ | \alpha> = |
\alpha_d> $ which contain the 2-body $^3S_1-^3D_1$ states we  
extract the deuteron pole. Thus we define
\begin{eqnarray}
t_{l_{\alpha} l_{\bar \alpha}}^{s_{\alpha} j_{\alpha} t_{\alpha}}
(p,p~';E(q)) \equiv \frac{\hat t_{l_{\alpha} l_{\bar
\alpha}}^{s_{\alpha} j_{\alpha} t_{\alpha}} (p,p~';E(q))}{E +i\epsilon  
- \frac{3}{4m} q^2 - \epsilon_d}
\label{eq5}
\end{eqnarray}
for the deuteron quantum numbers $s_{\alpha} = j_{\alpha }= 1,
 t_{\alpha }=0, l_{\alpha}, l_{\bar \alpha}=0,2$ and keep $t$ as
it is otherwise. That pole property obviously carries over to the
$ T$-amplitude and we define just for the $ | \alpha> = |
\alpha_d> $ channels
\begin{eqnarray}
\label{eq6}
\langle p q \alpha \vert T  \rangle = \frac{\langle p q
\alpha \vert  \hat  T  \rangle}{E +i \epsilon  -
\frac{3}{4m} q^2 - \epsilon_d} ~.
\end{eqnarray}
Since the energy $E$ of the 3N system is determined by the incoming
neutron energy $E_{cm}$: $E=E_{cm} + \epsilon_d \equiv \frac {3} {4m}
q_0^2 + \epsilon_d$, the deuteron pole occurs at $q=q_0$. 

When beside the deuteron an additional bound state exists in some
2-nucleon partial wave state one needs to extract in  channels  $ | \alpha>$
which contain that 2-nucleon state the corresponding pole of the
t-matrix by performing  the same procedure as for the deuteron. 
Let us assume that this state is a bound state of two neutrons in the 
$^1S_0$ state with the wave function $\phi_{nn}$ and the binding
energy $\epsilon_{nn}$,  and let us denote by 
 $| \Phi_{1,2} > \equiv \vert {\vec {\bar {q_0}}}, \phi_{nn} >$ the 
two body channel build on such dineutron,
 from which or to which
different reactions can be initiated. 

In the channels $ | \alpha> = |\alpha_{^1S_0}> $ which contain the
$^1S_0$ dineutron  we  define
\begin{eqnarray}
t_{l_{\alpha} l_{\bar \alpha}}^{s_{\alpha} j_{\alpha} t_{\alpha}}
(p,p~';E(q)) \equiv \frac{\hat t_{l_{\alpha} l_{\bar
\alpha}}^{s_{\alpha} j_{\alpha} t_{\alpha}} (p,p~';E(q))}{E +i
\epsilon  - \frac{3}{4m} q^2 - \epsilon_{nn}} = 
\frac{\hat t_{l_{\alpha} l_{\bar
\alpha}}^{s_{\alpha} j_{\alpha} t_{\alpha}} (p,p~';E(q))}
{\frac{3}{4m}({\bar q_0}^2- q^2) +i \epsilon}
\label{eq7}
\end{eqnarray}
and the dineutron pole occurs at $q=\bar q_0 = \sqrt{q_0^2 + \frac
  {4m} {3}(\epsilon_d - \epsilon_{nn})}$. 
Again that pole property carries over to the 
$ T$-amplitude and we define  for the $ | \alpha> = |
\alpha_{^1S_0}> $ channels the amplitude $\langle p q
\alpha \vert  \hat  T \rangle$  similarily to 
Eq.~(\ref{eq6}). The numerical treatment of that new pole follows
the treatment of the deuteron pole~\cite{physrep96} and it requires
the set of q-points which, in addition to $q=q_0$ needed for
numerical treatment of the deuteron pole, contains also $q=\bar
q_0$ point. 
 Since the dineutron occurs in the neutron-neutron 
  $^1S_0$ state it implies charge
 independence breaking and the resulting difference between
 $^1S_0$ nn and np interactions causes that the proper treatment of the nd
 reactions requires inclusion of the total 3N system isospin component
 $T=3/2$ for channels $\alpha$ containing $^1S_0$~\cite{witcib}.

The existence of $^1S_0$ dineutron  increases number of possible reactions
with three nucleons what in consequence makes  that 
  the unitary relation have to be generalized to include those
additional processes. It has the form
\begin{eqnarray}
& &   < \Phi_{ 1,a}| U | \Phi_{1,a'}>^* - < \Phi_{1,a'}| U
  |\Phi_{1,a}> \cr
&  = & 2 \pi i \sum_{b=1,2} \int d^3 q < \Phi_{ \vec q ,b}|  U | \Phi_{1,a'}>^*
 \delta ( E_{\vec q} ^b - E_{ \vec q}) < \Phi_{ \vec q, b} | U | \Phi_{ 1,a} >\cr
 & + &  2\pi i  /6 \int d^3 p d^3 q < \Phi_0|  U_0 | \Phi_{ 1,a'}>^*
 \delta (E_{pq} - E_{\vec q} ) < \Phi_0 | U_0 | \Phi_{1,a} >
\label{eq8}
\end{eqnarray}
with $a = 1$ and $2$ for the  deuteron  and dineutron channels, respectively. 
 One can choose  $ a'=1$  or $ a'=2 $   and $ a = 1$  or $a=2$.
For $a=a'=1$ this leads on the left side to the 
forward scattering amplitude and on the right to the  total cross
section. 
The energies $E_{\vec q}^b = E_{\vec q} + E_b$ are given by the
binding energies of the deuteron $E_1 = \epsilon_d$ or 
dineutron $E_2 = \epsilon_{nn}$. 

The angular distribution for the process $n + d \to p + dineutron$
is given by the transition amplitude $<\Phi_{1,2}|U|\Phi_{1,1}>$
\begin{eqnarray}
\frac {d\sigma} {d\Omega}({n+d \to p + dineutron}) &=& (\frac {2m}  {3})^2 
 (2\pi)^4 \frac {{\bar q}_0} {q_0} \sum_{m_p m_n m_d}
|\left\langle {\Phi _{1,2}} \right|U\left| {\Phi _{1,1} }
\right\rangle|^2 ~,
\label{eq9}
\end{eqnarray}
where $PG_0^{-1}$ and  $PT$   contributions to $U$ are given by
\begin{eqnarray}
\left\langle {\Phi _{1,2}} \right| PG_0^{-1} \left| {\Phi _{1,1} }
\right\rangle &=& \left\langle {\phi_{nn}}, m_p, {\vec
  {\bar q}_0 }\right| PG_0^{-1} \left| {\phi_d, m_n, m_d,
  \vec q_0 || \hat z} \right\rangle  \cr
&=&  \frac {2} {\sqrt{4\pi}} 
 [ \epsilon_d - \frac{1}{m}( \frac {1} {4} q_0^{2} + \bar q_0^2 + 
\vec q_0 \cdot {\vec {\bar q}_0} )]  
( \frac {1} {2}  \frac{1}{2} 1 | - \frac {1} {2},  - \frac {1} {2}, -1)
( \frac {1} {2}  \frac{1}{2} 0 | - \frac {1} {2}, \frac {1} {2}, 0) \cr
&& \phi_{nn} ( | \vec q_0 + \frac{1}{2} \vec {\bar q}_0 |) 
 \sum_{ l=0,2} ( l 11| m_d + m_n - m_p, -m_n + m_p, m_d) \cr
&& ( \frac {1} {2}  \frac{1}{2} 1 | -m_n, m_p, -m_n + m_p)
( \frac {1} {2}  \frac{1}{2} 0 | m_n, -m_n, 0) \cr
   & & \phi_l^d ( | \frac{1}{2} \vec q_0 + \vec {\bar q}_0 |) 
Y_{ l, m_d + m_n  - m_p} (   { \frac{1}{2} \vec q_0  + \vec {\bar
    q}_0 }) 
\label{eq11}
\end{eqnarray}
and
\begin{eqnarray}
\left\langle {\Phi _{1,2}} | P |T \right\rangle 
 &=& \left\langle {\phi_{nn}}, m_p, {\vec
  {\bar q}_0 } | P | T  \right\rangle  \cr
&=&  \sum_{J^{\pi} M} \sum_{\alpha' \alpha_0} \delta_{I_0J}
\delta_{l_00} \delta_{s_00} \delta_{j_00} 
 ( \lambda_0 \frac{1}{2} I | M-\mu', \mu', M) 
 ( 1  \frac{1}{2} T_0 | -1, \frac {1} {2}, -\frac {1} {2} )
 Y_{ \lambda_0, M-\mu'} ( \hat {\bar q}_0) \cr 
&&  \int_{0}^{\infty} q'^2dq'  \int_{-1}^1 dx ~
    \phi_{nn}( \pi_1) \frac {G_{ \alpha_0,\alpha'} ( \bar q_0,q',x)} 
 {\pi_1^{l_0} \pi_2^{l_{\alpha'}} }
\left\langle \pi_2, q', \alpha' | T \right\rangle ~,
\label{eq10}
\end{eqnarray}
with
\begin{eqnarray}
\pi_1 &=& \sqrt{q'^2 + \frac {1} {4} \bar q_0^2 + q'\bar q_0 x}~, \cr
\pi_2 &=& \sqrt{\bar q_0^2 + \frac {1} {4} q'^2 + q'\bar q_0 x}~. 
\label{eq12}
\end{eqnarray}
It was assumed that the relative neutron-deuteron momentum $\vec q_0$
in the incoming channel is directed along the z-axis. The convention for
isospin projections is that for the neutron it is $-\frac {1} {2}$ while
for  the proton $ +\frac {1} {2}$. In Eq.~(\ref{eq10}) channels
$\alpha_0$ contain the dineutron two-nucleon subsystem quantum numbers 
with isospin $t_0=1$ and its projection $\nu_{t_0}=-1$ and the total
isospin $T_0$ of the 3N system for these channels is 
$T_0= \frac {1} {2}$ or  $T_0= \frac {3} {2}$. The geometrical
coefficient $G_{ \alpha_0,\alpha'} ( \bar q_0,q',x)$ stems from the
 matrix elements of the permutation operator $P$~\cite{physrep96}.

\section{ Results}
\label{sec2}

In the following we will present and compare  to the available nd data
 the theoretical predictions 
for cross sections in elastic nd scattering and breakup assuming
different  $^1S_0$ nn force. We take the
CD~Bonn~\cite{cdbonn} 
potential as the NN interaction and multiplying its $^1S_0$ nn
component by a factor $\lambda$  generate a number of $^1S_0$ nn forces among
which some provide binding of two neutrons. In Table~\ref{tab1} we
show values of the nn scattering length $a_{nn}$, the effective range
parameter $r_{eff}$ and the dineutron binding energy $\epsilon_{nn}$ for
a number of $\lambda$ values. Changing $\lambda$ from $0.9$ to $1.5$
leads to nn $^1S_0$ force with different, negative as well as
positive, values of the scattering length. 
 In order to see if conclusions depend from a particular $^1S_0$ nn  potential
 used and from the method applied to generate the nn bound state, 
 we performed also 
calculations with a chiral
 NN potential in next-to-leading-order (NLO) of chiral
 expansion~\cite{epel} 
 adjusting its  $^1S_0$ nn low energy constants  to get a 
 dineutron with given binding energy.

\subsection{Total cross sections} 
\label{sec3}

\begin{figure}
\includegraphics[scale=0.8]{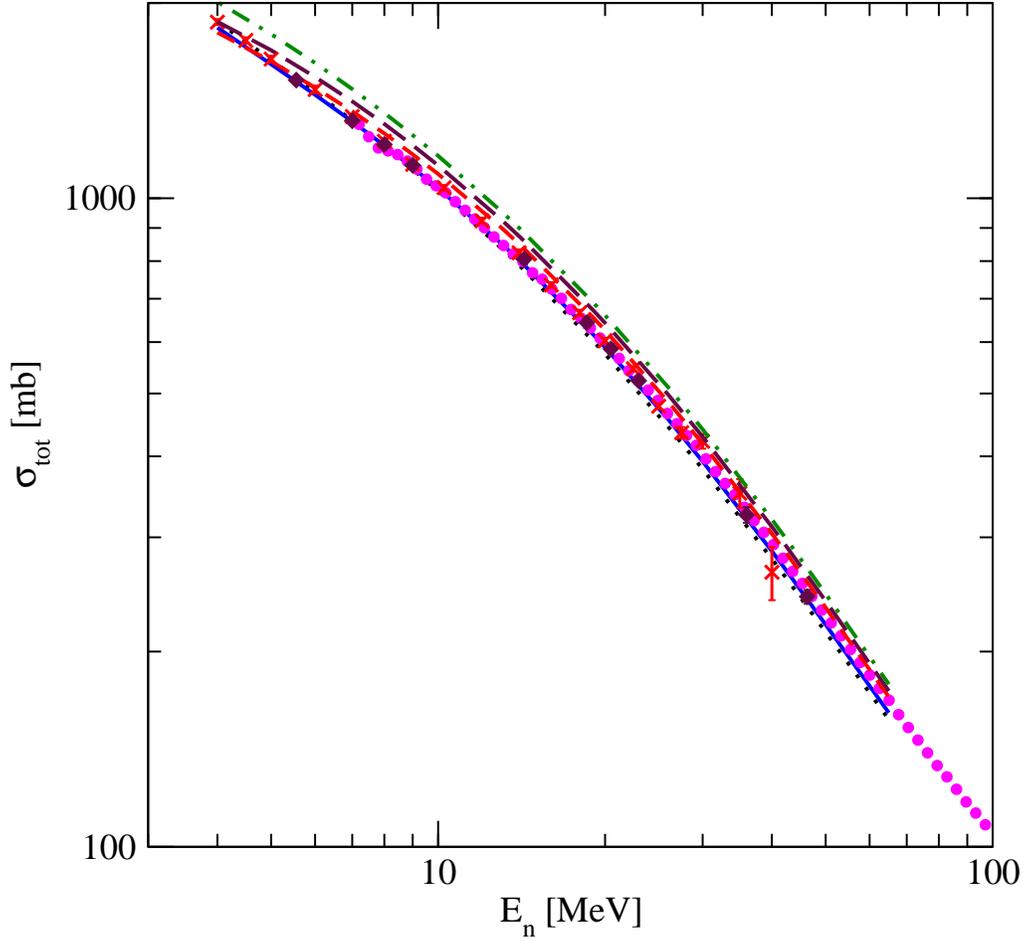}
\caption{(color online) The total cross section for the neutron-deuteron
  interaction as a function of the neutron lab. energy. 
 Different lines  show sensitivity of the total 
cross section to the changes of the nn $^1S_0$ force component. Those
changes were induced by multiplying the $^1S_0$ nn matrix element of the CD
Bonn potential by a factor $\lambda$. The solid (blue) line is the full
result based on the original CD Bonn potential 
 ($\lambda=1.0$) and all partial
waves with 2N total angular momenta up to $j_{max}=3$ included. The
(black) dotted, (red) short-dashed, 
(maroon) long-dashed, and (green) dashed-double-dotted  lines correspond to
$\lambda=0.9$, $1.21$, $1.3$, and $1.4$, respectively.  
 The (magenta) circles, (red) x-es,  and (maroon) diamonds 
are nd data of Ref. \cite{LANL}, \cite{Schwarz}, and \cite{Seagrave},
  respectively. 
}
\label{fig1}
\end{figure}

The results for the nd total cross sections are shown in 
Fig.~\ref{fig1} and, for a number of energies, they are also presented
  in Table~\ref{tab2}. The theoretical predictions obtained
with different nn $^1S_0$  forces are compared to  numerous data
taken  over many years. Up to $\approx 100$~MeV
there is a nice agreement between all data, especially very precise one of
Ref.~\cite{LANL}, and  theory based on the CD~Bonn potential. When instead
 of the original CD~Bonn $^1S_0$ nn force the modified interaction with
factor $\lambda=0.9$  is taken the resulting cross
sections seem to be  not excluded by the total cross section data. 
 For  $\lambda=1.21$, with the dineutron binding energy
 $\epsilon_{nn}=-144$~keV, the predicted total cross sections 
 for energies up to $\approx 10$~MeV differ
 from the  data by about three standard deviations. At higher energies 
they clearly lie outside three standard deviations from the data.  
Increasing the factor $\lambda$ to $1.3$ or $1.4$ leads to total cross
section values strongly overestimating the data.

\begin{figure}
\includegraphics[scale=0.8]{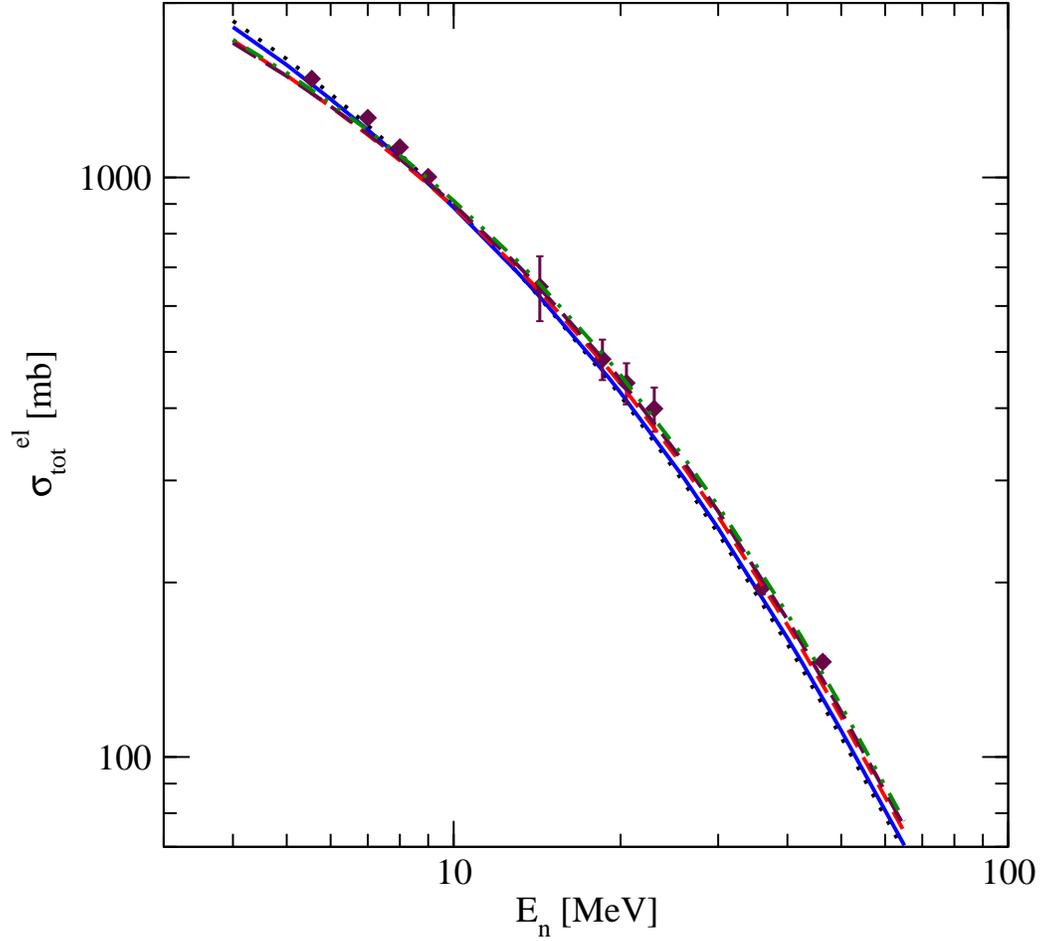}
\caption{(color online) 
The total neutron-deuteron elastic scattering cross section 
 as a function of the neutron lab. energy. 
 Different lines  show sensitivity of that 
cross section to the changes of the nn $^1S_0$ force component. For
their description see Fig.\ref{fig1}. 
 The (maroon) diamonds 
are nd data of Ref. \cite{Seagrave}. 
}
\label{fig2}
\end{figure}

\begin{figure}
\includegraphics[scale=0.8]{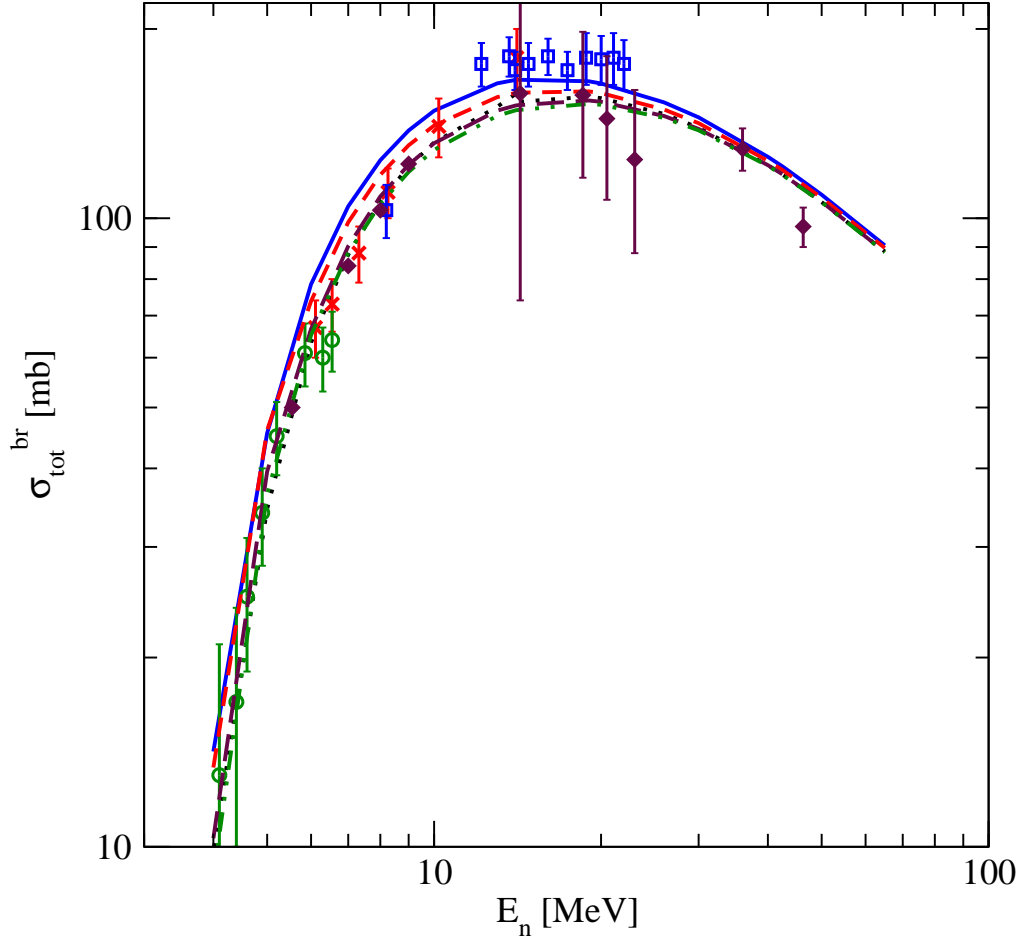}
\caption{(color online) 
The total neutron-deuteron breakup cross section 
 as a function of the neutron lab. energy. 
 Different lines  show sensitivity of that 
cross section to the changes of the nn $^1S_0$ force component. For
their description see Fig.\ref{fig1}. 
 The (red) crosses, (green) circles, (blue) squares, and (maroon)
 diamonds are nd data of Ref. \cite{Catron}, \cite{Holmberg},
 \cite{Pauletta} and \cite{Seagrave}, 
  respectively. 
}
\label{fig3}
\end{figure}

In Figs.~\ref{fig2} and \ref{fig3} we compare theoretical predictions
for the total elastic scattering and breakup cross sections,
respectively, with the corresponding data. 
 For the elastic
scattering component of the total cross section (see Fig.~\ref{fig2}), 
at energies up to about $E_n \approx 20$~MeV, 
 the theoretical predictions with  different nn $^1S_0$  forces 
 are  close to each 
other and they agree with the data.
At energies above $E_n \approx 20$~MeV cross sections for $\lambda > 1$
start to deviate from standard CD~Bonn and $\lambda=0.9$ values and
the data seems to prefer larger values of $\lambda$. 

For the total breakup cross sections (see Fig.~\ref{fig3}) 
 the data seems to be compatible
with all theoretical predicitons with exception of data from
Ref.~\cite{Pauletta}. That data set taken in the region of energies
$12$~MeV $ < E_n < 22$~MeV  clearly advocates  the CD~Bonn potential 
predicitons However, it does not exclude definitly values of cross
sections obtained with $\lambda=1.21$. 

At low energies the nd interaction is parametrized by the doublet, 
$^2a_{nd}$, and quartet, $^4a_{nd}$, scattering lengths. While $^2a_{nd}$
is strongly influenced by a 3NF the $^4a_{nd}$ is
practically insensitive to such an interaction~\cite{witndscat}. 
 In Table~\ref{tab3} we
show how these scattering lengths change with modification of the
$^1S_0$ nn CD~Bonn potential. While the doublet scattering length
drastically changes with $\lambda$ the quartet scattering length
practicaly remains constant under such modifications of the  $^1S_0$
nn force staying close to the experimental value 
 $^4a_{nd}=(6.35 \pm 0.02)$~fm~\cite{Dilg}.

\subsection{Elastic neutron-deuteron scattering}
\label{sec4}

\begin{figure}
\includegraphics[scale=0.6]{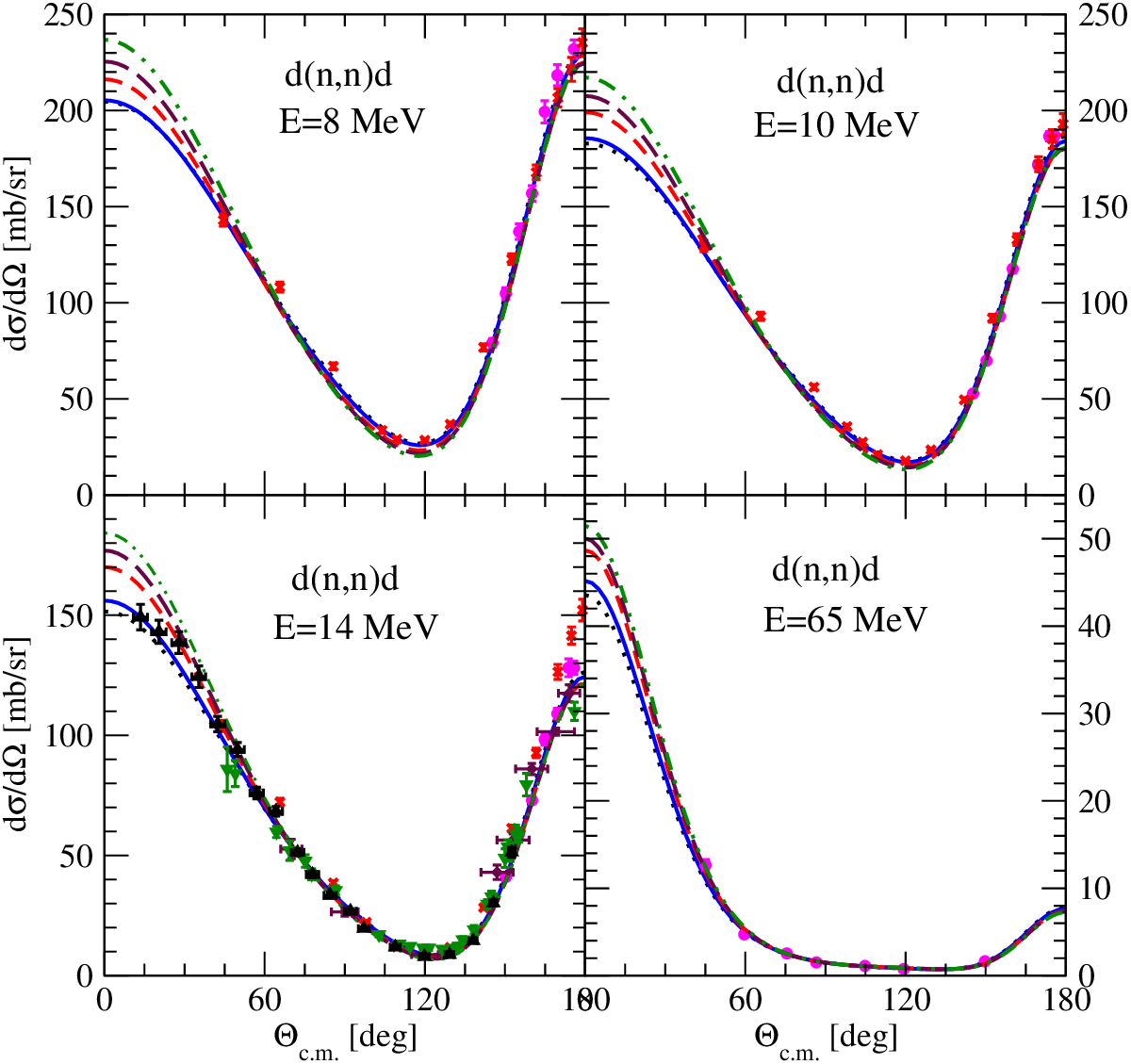}
\caption{(color online) The  neutron-deuteron elastic scattering 
angular distributions $d\sigma/d\Omega$ at a number of incoming
neutron lab. energies. 
 Different lines  show sensitivity 
 to the changes of the nn $^1S_0$ force component. For
their description see Fig.\ref{fig1}. 
At $E_n=8, 10$ and $14$~MeV  the (magenta) circles and (red) x-es 
are nd data of Ref. \cite{TUNL_sig} and \cite{Schwarz}, respectively.  
At $E_n=14$~MeV  the (maroon) stars, (green) triangle-down,  and
(black) triangle-up  
are nd data of Ref. \cite{Seagrave}, \cite{Allred}, and
\cite{Berick},  respectively.  
At $E_n=65$~MeV  the (blue) circles 
are $E_n=66$~MeV nd data of Ref. \cite{PSI}.
}
\label{fig4}
\end{figure}

The nd elastic scattering angular distributions are shown in
Fig.~\ref{fig4}. At c.m. scattering angles $\Theta_{c.m.} > 45^o$
different theories practically overlapp and agree with the data for
all four energies shown. Such behaviour is not suprising since at
backward angles the exchange term $PG_0^{-1}$, given by the deuteron
 wave function, dominates the elastic scattering transition
amplitude. The  properties of the  nn
$^1S_0$ interaction should  play decisive role at forward angles. 
 Indeed, at forward angles below 
$\Theta_{c.m.} < 45^o$  differences between theoretical predictions based on
 different nn $^1S_0$  forces start to appear and they increase with
 deacreasing angle. However, in the forward angular region the nd
 elastic scattering cross section data  are
 lacking with exception of $E_n= 14.1$~MeV where 5 data points fall
 into that region of angles. While two data points at smallest angles support
 the  CD~Bonn cross sections three other at greater angles prefer the 
larger values of  $\lambda$. The precise nd elastic scattering data 
at forward angles are required to decide if stronger nn $^1S_0$ force
is allowed.

\subsection{Breakup}
\label{sec5}

Among numerous kinematically complete breakup configurations 
  the largest discrepancies between  theory and data
have been found for the nn QFS and SST
geometries. For these configurations the theoretical cross sections
are insensitive to the underlying dynamics and they do not change when
applying different realistic NN potentials and combining them with 
 available 3NF's. Also when
instead of the nn QFS one compares the theory with the only one
available np QFS data set of the nd breakup~\cite{siepe2} a nice
agreement is found. QFS and SST configurations are dominated  at low energies 
 by the $^1S_0$ and $^3S_1$ NN force 
components~\cite{wit1} which practically saturate the QFS and SST 
cross section at low energies~\cite{wit1,wit2}.  
It would suggests that it is the nn $^1S_0$
force which is probably responsible for large discrepancy between data
and theory. 

\begin{figure}
\includegraphics[scale=0.8]{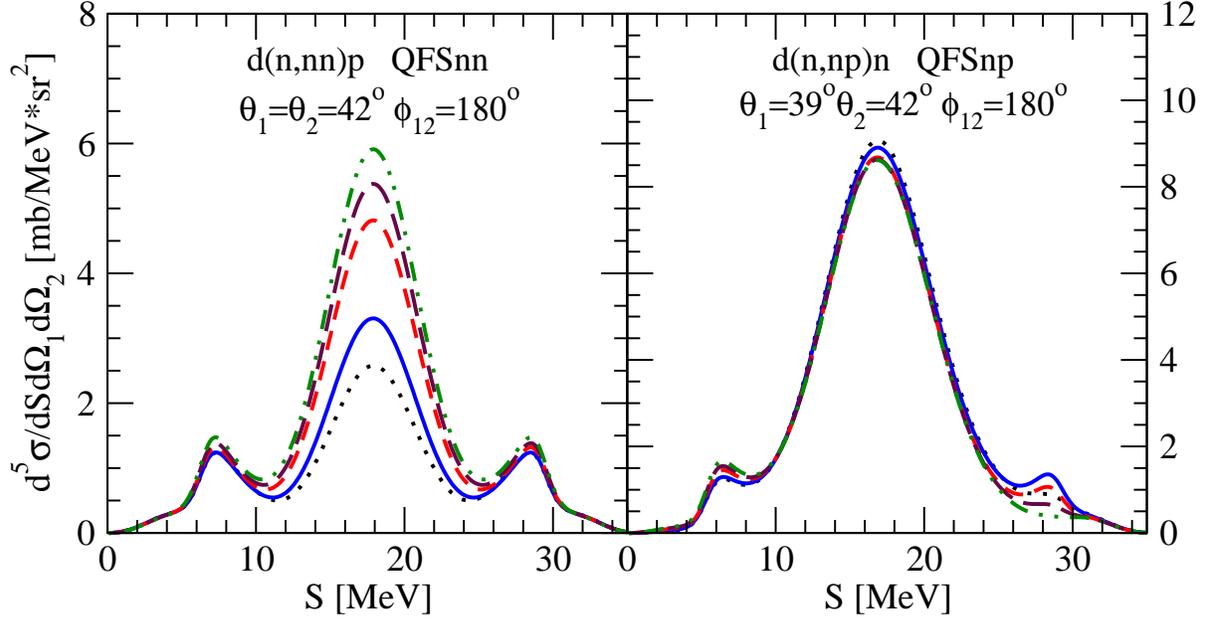} 
\caption{(color online) The cross section
  $d^5\sigma/d\Omega_1d\Omega_2dS$ 
  as a function of the S-curve arc-length in 
 the $E_n^{lab}=26$~MeV nd breakup reaction $d(n,nn)p$ for the QFS nn
 (left) and np (right) kinematically  
complete configurations of Ref.~\cite{siepe2}. For description of
lines see Fig.~\ref{fig1}.}
\label{fig5}
\end{figure}

For QFS configurations we show  in Fig.~\ref{fig5} the sensitivity 
of the nn and np QFS configurations to the underlying nn
$^1S_0$ force. As expected, changes of that force cause drastic
modifications of the nn QFS cross section leaving the np QFS
practically without modifications. As was shown in \cite{wit2} 
 responsible for such drastic modifications of the
 nn QFS cross section are changes of the effective range parameter
 induced by factor $\lambda$. The changes in the nn scattering
length practically leave the nn QFS cross sections without modifications. 

\begin{figure}
\includegraphics[scale=0.7]{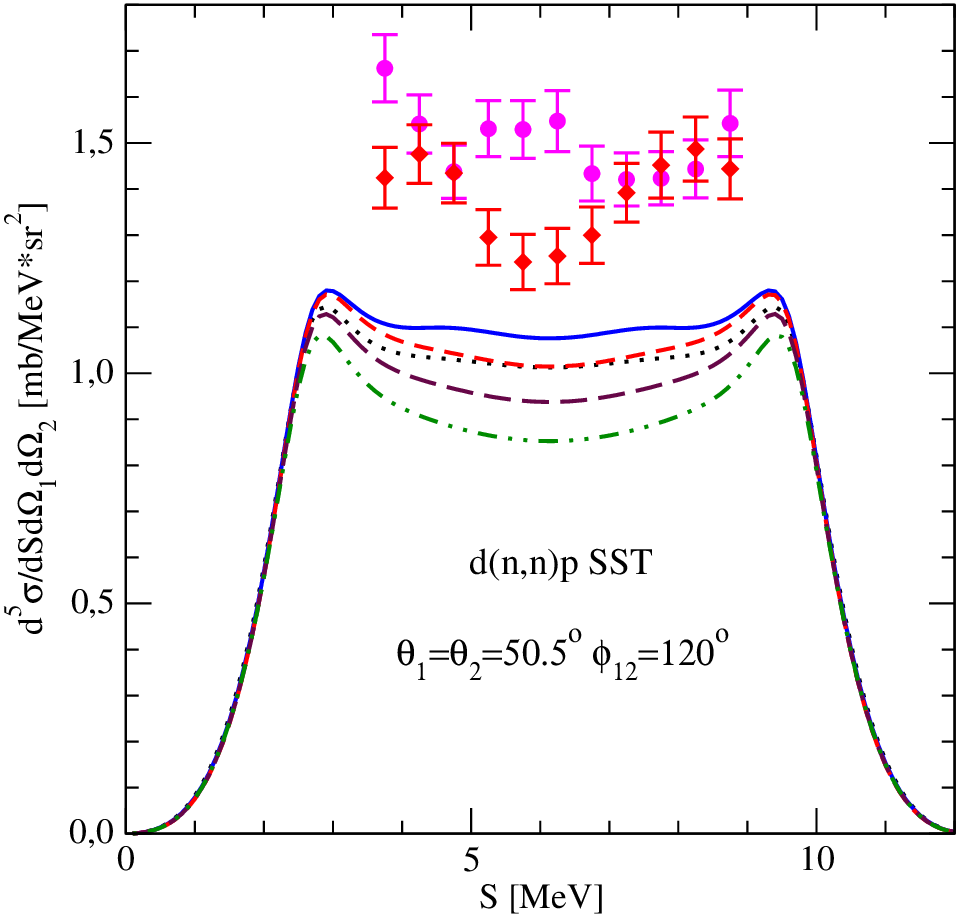}
\caption{(color online) The cross section
  $d^5\sigma/d\Omega_1d\Omega_2dS$ 
  as a function of the S-curve arc-length 
 in the $E_n^{lab}=13$~MeV nd breakup reaction $d(n,nn)p$  
for SST  configuration with the lab. angles of two
detected neutrons $\theta_1=\theta_2=52.8^o$ and $\phi_{12}=180^o$. 
 For description of lines see Fig.~\ref{fig1}.
 The (magenta) solid dots and (red) x-ses
are nd data of Ref. \cite{tunl1,tunl2} and \cite{strate1}, respectively.}
\label{fig6}
\end{figure}

In contrast to the nn QFS  the SST geometry is more stable  
against changes of the $^1S_0$ nn force. As shown in Fig.~\ref{fig6}
changing the  factor $\lambda$ does not bring theory closer to the data. 
While $\lambda=0.9$ provides smaller SST cross sections than  the CD~Bonn
potential, taking factor $\lambda > 1$ and increasing it so that 
 dineutron is formed,  leads to  cross sections which again are below
the CD~Bonn potential predicitons. Therefore by modifications of the $^1S_0$ nn
force it is not possible to explain the large discrepancy for SST. Since it is
unprobable that the deuteron properties are so badly known that the
$^3S_1-^3D_1$ NN force component would require modification, the source
 for that  disagreement must be sought elsewhere. 
 One possibility
 could be the indirect influence by the dineutron some breakup
 configurations by  contributing in specific regions of a 
 phase-space to the breakup background. 

\begin{figure}
\includegraphics[scale=0.84]{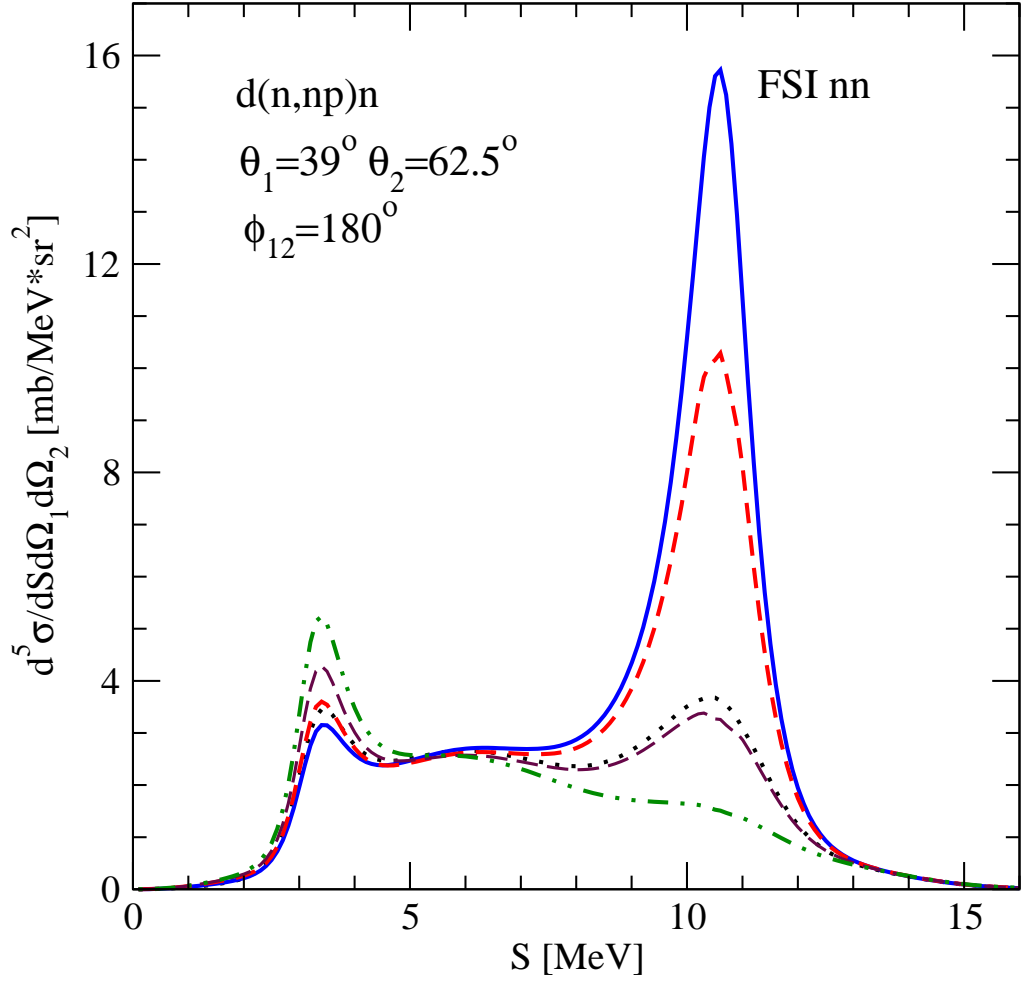}
\caption{(color online) The cross section $d^5\sigma/d\Omega_1d\Omega_2dS$
for the $E_n^{lab}=13$~MeV nd breakup reaction $d(n,np)n$  as a function of
the S-curve length for nn FSI configuration. 
 For description of lines see Fig.~\ref{fig1}.}
\label{fig7}
\end{figure}

For the $^1S_0$ nn force which allows a dineutron the nn scattering length
 becomes  positive. It should have drastic influence on the
 nn final-state-interaction (FSI) of the nd breakup, where the two
 outgoing neutrons, having the same momenta,  are strongly 
interacting in the $^1S_0$ state. We show in Fig.~\ref{fig7} the
changes in the FSI peak when the nn scattering length $a_{nn}$ changes 
from negative to positive values. For the same magnitude of 
$a_{nn}$ the nn FSI cross section is strongly diminished 
for the positive sign of
$a_{nn}$. The question arises if the existing nn FSI cross section data can be
understood when dineutron exists ?

\begin{figure}
\includegraphics[scale=0.75]{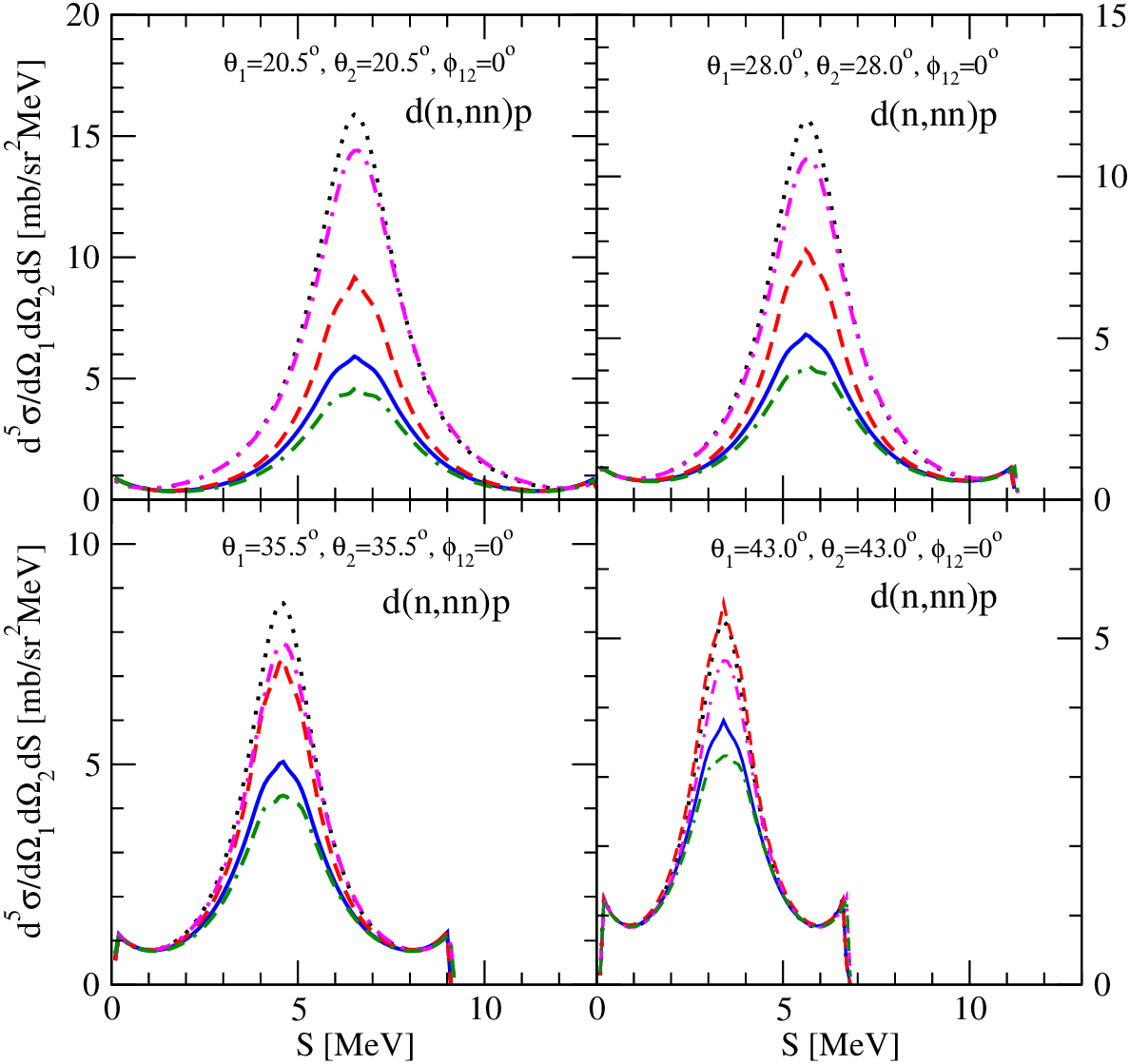} 
\caption{(color online) The cross section
  $d^5\sigma/d\Omega_1d\Omega_2dS$ 
  as a function of the S-curve arc-length in 
 the $E_n^{lab}=13$~MeV nd breakup reaction $d(n,nn)p$ for 4 FSI nn
 geometries. Different lines  show sensitivity of the 
cross section to the changes of the nn $^1S_0$ force component. Those
changes were induced for the dotted (black), (red) dashed and  
 (blue) solid  lines by multiplying the $^1S_0$ nn matrix element of the CD
Bonn potential by a factor $\lambda$. The dotted (black) line is the full
result based on the original CD Bonn potential 
 ($\lambda=1.0$) and all partial
waves with 2N total angular momenta up to $j_{max}=3$ included. The
(red) dashed and  (blue) solid  lines correspond to
$\lambda=1.19$ and $1.21$,  respectively.
 The (magenta) dashed-dotted and (green) double-dashed-dotted  lines
 show results of
Faddeev calculations based on NLO chiral perturbation theory potential 
 and all partial waves with 2N total angular momenta up to 
 $j_{max} = 3$  included. 
 They differ in the nn $^1S_0$ force which for the (magenta)
 dashed-dotted line was obtained with the constants $C_1(^1S_0)=1.0$ 
 and $C_2(^1S_0)=1.0$ (original NLO potential, see text for
 explanation) leading to
 $a_{nn} = -17.6$~fm and   $r_{eff} = 2.75$~fm. 
 For the (green) double-dashed-dotted line  the constants $C_1(^1S_0)=1.50$ 
 and $C_2(^1S_0)=1.29415$, what results in 
 $a_{nn} = +17.5$~fm and   $r_{eff} = 2.41$~fm.}
\label{fig8}
\end{figure}

To answer that question we show in Fig.~\ref{fig8} cross sections for
4 kinematically complete nn FSI configurations for which data have been
taken  and analyzed in Ref.~\cite{tunlfsinn} with the aim to extract the
neutron-neutron scattering length. The consistent values of 
 $a_{nn}$ have been found in each of those 4
configurations with the average value of $a_{nn} = −18.7 \pm 0.7$~fm. 
 As can be seen in Fig.~\ref{fig8}, again changing 
 $a_{nn}$ to positive values diminishes significantly the nn FSI cross
sections. Comparing cross sections obtained
with $\lambda=1.19$ and $\lambda=1.21$ to the CD~Bonn potential values
  clearly demonstrates that no theoretical analysis of
  \cite{tunlfsinn} data, when performed with
positive values of $a_{nn}$, would provide consistent values for the nn
scattering length in those 4 geometries. While the analysis of
$\theta_1=\theta_2=43^o$ configuration would probably provide
$a_{nn}=+21.69$~fm, a similar analysis of configurations at smaller 
$\theta_1=\theta_2$ would provide distinctly larger positive nn scattering
lengths. 

\begin{figure}
\includegraphics[scale=0.8]{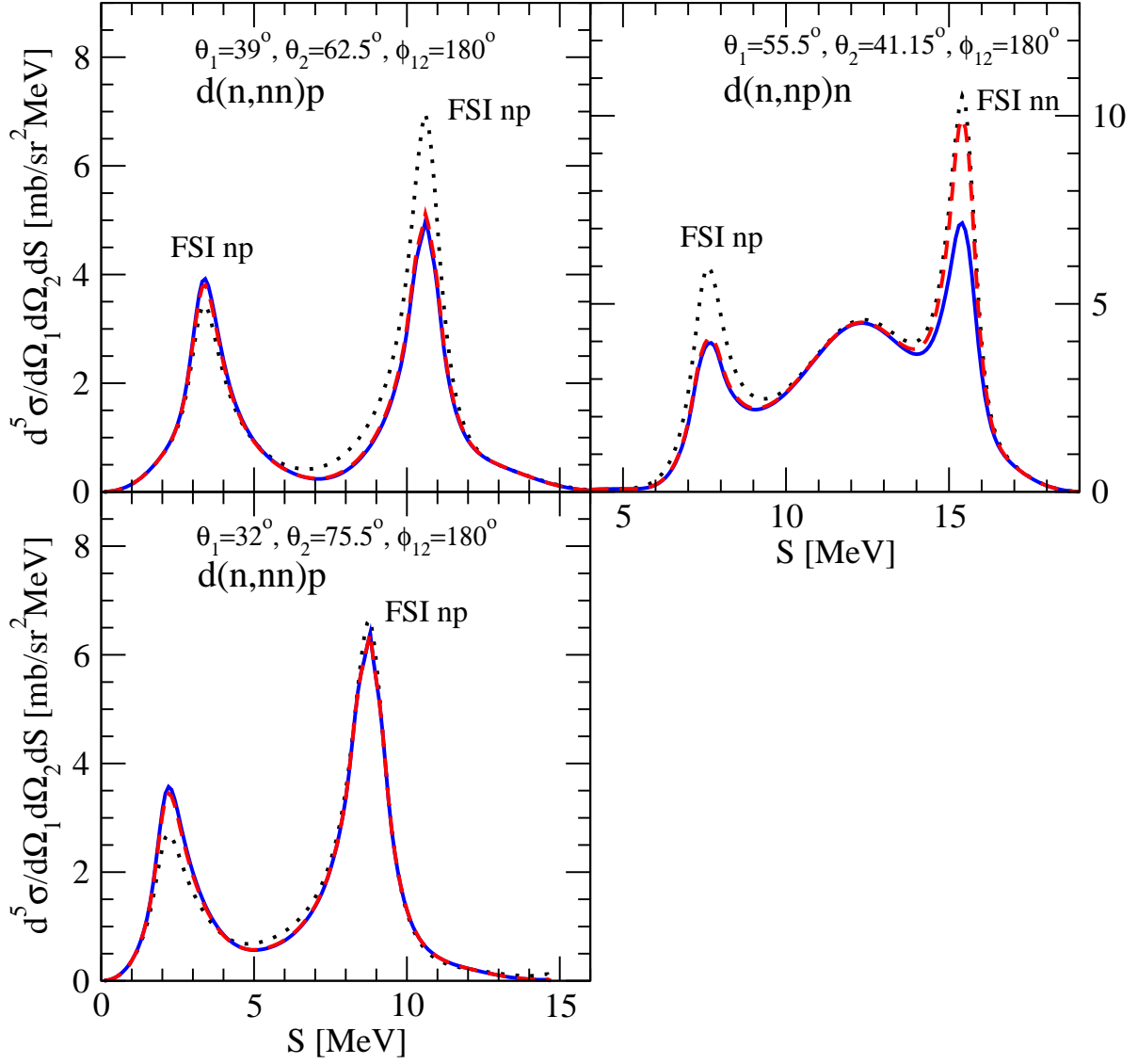} 
\caption{(color online) The cross section
  $d^5\sigma/d\Omega_1d\Omega_2dS$ 
  as a function of the S-curve arc-length in 
 the $E_n^{lab}=13$~MeV nd breakup reaction $d(n,nn)p$ for 3 FSI 
 geometries.   Different lines  show sensitivity of that 
cross section to the changes of the nn $^1S_0$ force component. For
their description see Fig.\ref{fig8}.} 
\label{fig9}
\end{figure}

In Fig.~\ref{fig9} we show further 3 FSI configurations for which
data have been taken. For $d(n,nn)p$ complete breakup the data of
Ref.~\cite{tunl2} support the CD~Bonn potential cross section predictions. 
Each of 2 complete configurations in that reaction shown in
Fig.~\ref{fig9} contain two np FSI peaks. The theoretical
analysis of these np FSI peaks, if performed with positive values of
$a_{nn}$, would provide different 
 values for the neutron-proton scattering length $a_{np}$, which in
 addition would be inconsistent with well
known $a_{np}$ experimental value.

In Fig.~\ref{fig9} we show also the configuration for $d(n,np)n$ breakup
 in which data have been taken and analyzed in Ref.~\cite{Huhn}.  That
geometry contains both np and nn FSI peaks. Again, the analysis of the
np FSI peak, if performed with positive $a_{nn}$, would 
provide too large magnitude for $a_{np}$.  

To see how our conclusions depend on the NN potential used and on the
 method applied to modify the $^1S_0$ nn force we present in
Fig.~\ref{fig8} also cross sections obtained with
next-to-leading-order (NLO) chiral perturbation theory 
 potential of Ref.~\cite{epel}, 
 including in calculations all np
and nn forces up to the total angular momentum $j_{max} = 3$ in the
two-nucleon subsystem. The $^1S_0$ component of that interaction
is composed of the one- and two-pion exchange terms and
contact interactions parametrized by two parameters $\tilde C_{^1S_0}$ and
 $C_{^1S_0}$
\begin{eqnarray}
V (^1S_0) = \tilde C_{^1S_0} + C_{^1S_0} (p^2 + p'^2) ~.
\end{eqnarray}
Standard values are $\tilde C_{^1S_0} = -0.155 737 4 \times  10^4$~GeV$^{−2}$ 
and $C_{^1S_0} = 1.507 522 0 \times 10^4$~GeV$^{−4}$ for cutoff combinations
$ \{\Lambda, \tilde \Lambda \} = \{ 450$~MeV$, 500$~MeV$ \}$~\cite{epel}. 
By multiplying $\tilde C_{^1S_0}$ by a factor $C_2(^1S_0)$ and
$C_{^1S_0}$  by a factor $C_1(^1S_0)$, one can induce changes of the
nn $^1S_0$  interaction. In Fig.~\ref{fig8} we show two predictions
based on the  NLO potential with negative ($a_{nn}=-17.6$~fm - the (magenta) 
dashed-dotted line) and positive  ($a_{nn}=+17.5$~fm - the (green) 
 double-dashed-dotted  line) values of the neutron-neutron scattering
 length. Comparing them with different CD~Bonn potential predictions
 and taking into account differences between their $a_{nn}$ values it
 is clearly seen that both potentials and methods of changing $^1S_0$
 nn interaction lead to the same conclusions. 

\begin{figure}
\includegraphics[scale=0.84]{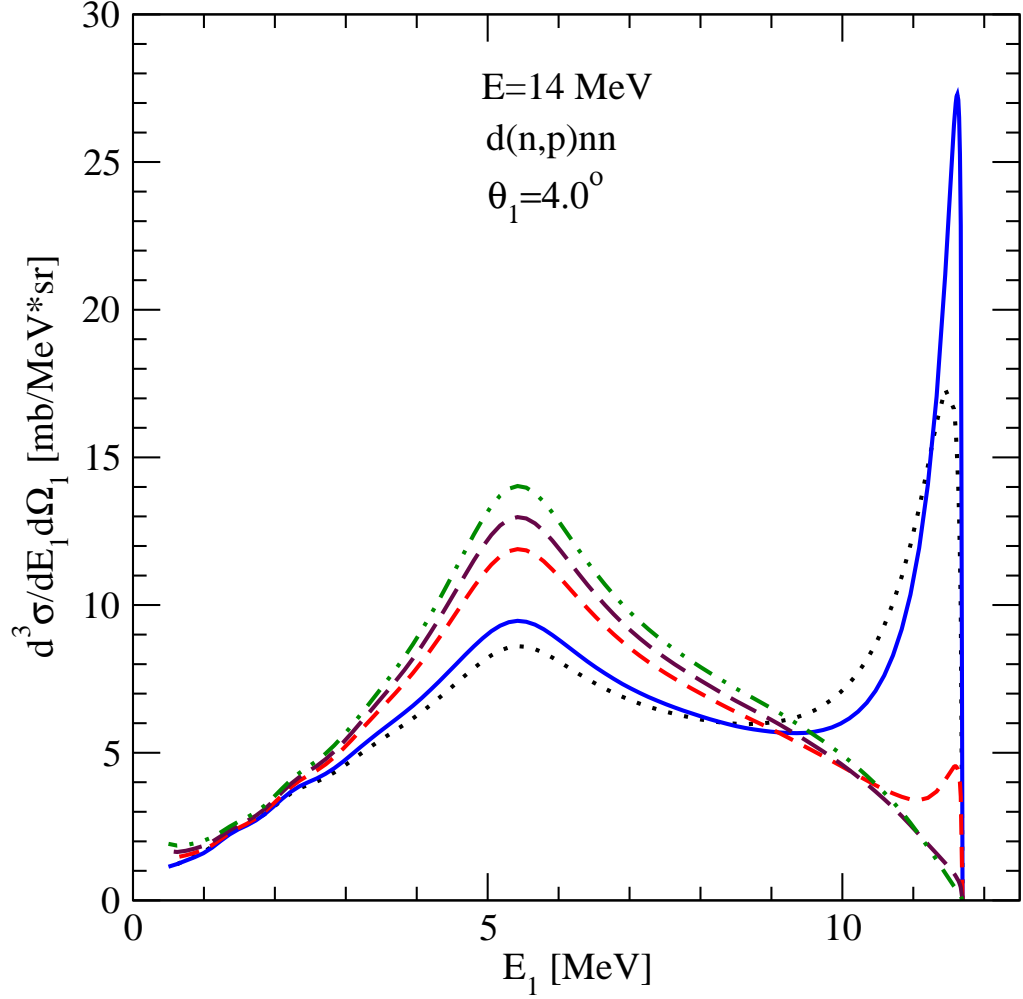}
\caption{(color online) The cross section $d^3\sigma/d\Omega_1dE_p$
for the $E_n^{lab}=14$~MeV uncomplete nd breakup reaction $d(n,p)nn$  
as a function of the outgoing proton lab. energy at the proton lab. angle 
 $\theta_1=4^o$.  For description of lines see Fig.~\ref{fig1}.}
\label{fig10}
\end{figure}

FSI region can also be investigated in the  uncomplete breakup 
measurement, in which   spectrum of the outgoing proton is measured 
 at given lab. angle. In Fig.~\ref{fig10} we show modifications of the
outgoing proton spectrum for $14$~MeV nd breakup at proton lab. angle 
$\theta = 4^o$. Here changing the sign of  $a_{nn}$ leads to
disappearing of the FSI peak. In addition , at lower  energies of the
outgoing proton the modification of $^1S_0$ nn force by factors 
$\lambda > 1$ significantly increases the uncomplete breakup cross section. 

The analysis of existing nd uncomplete breakup spectra performed in
\cite{tornow1,tornow2} indicated on the inconsistencies in the
experimental uncomplete nd breakup data  and revealed unexplained
differences of more than $25 \%$  in regions of the outgoing proton
energies where large number of different three-nucleon configurations
contribute to the cross section. 
The question arises if existence of the dineutron and corresponding
modification of the $^1S_0$ nn force can account for that and if the
clear FSI peaks appearing in the experimental outgoing proton spectra
provide evidence for existence of dineutron. In order to answer that
question a theoretical Monte Carlo analysis of experimental
spectra,  which would provide the angular distribution for the dineutron 
cross section, is
required. The resulting angular distribution should then be
compared to the theoretical angular distribution for $n+d \to p +
dineutron$ transition. However, in view of the results presented above
for the complete nn FSI configurations for which data are available, 
it seems highly unprobable that analysis of  incomplete spectra will
provide a clear signal for existence of  dineutron.

\subsection{Transition from the neutron-deuteron to the 
proton-dineutron channel}
\label{sec6}

\begin{figure}
\includegraphics[scale=0.6]{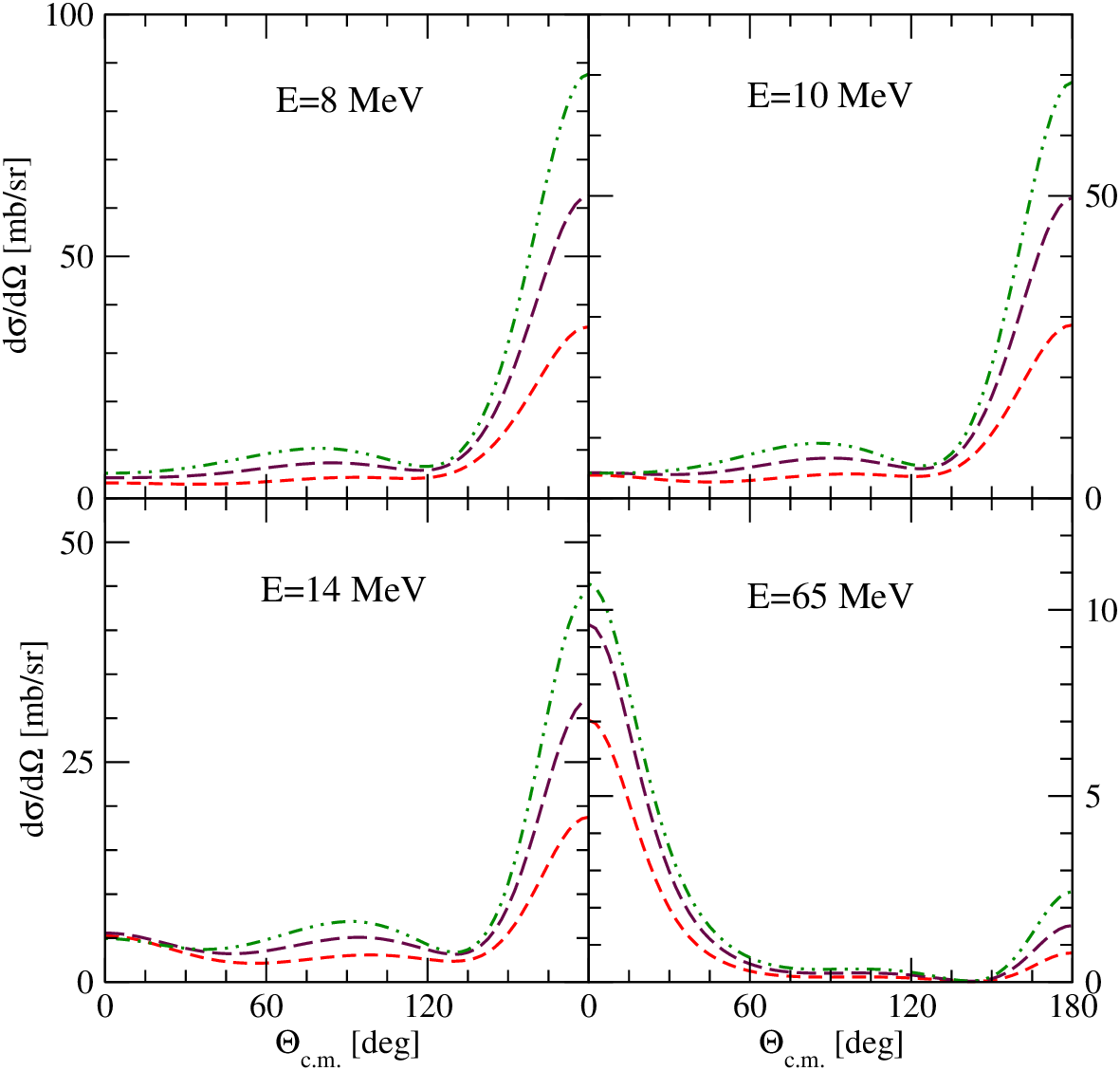}
\caption{(color online)  The angular distributions $d\sigma/d\Omega$ 
for $d(n,p)dineutron$ reaction 
at a number of incoming neutron lab. energies. The (red) short-dashed,
 (maroon) long-dashed, and (green) dashed-double-dotted lines 
correspond to the factor
$\lambda=1.21$, $1.3$, and $1.4$, respectively.  
}
\label{fig11}
\end{figure}

For values of factor $\lambda=1.21$, $1.3$ and $1.4$, which allow for the bound
$^1S_0$ state of two neutrons, the transition to the
proton-dineutron channel is possible. In Fig.~\ref{fig11} we show
 angular distributions for $n + d \to p + dineutron$ reaction. The
 cross sections for that reaction are by an  order of
magnitude smaller than for the nd elastic scattering,  with largest
cross sections at backward c.m. angles for low incoming neutron energies.

In view of the discrepancies found in the nd breakup reaction, especially
in SST configuration, it is interesting to consider in which
phase-space region that hypothetical dineutron state could mostly affect the
breakup configurations by contributing in an uncontrolled manner to the   
background. To answer that question again Monte Carlo
simulations of experimental conditions are required. 

\section{Summary and conclusions}
\label{sec7}

We have investigated how far  available  nd data allow for 
  hypothetical $^1S_0$ bound state of
two neutrons and if such dineutron can help to explain the  
discrepancies between theory and data found in some complete nd
breakup configurations. 
To this aim we extended our numerical momentum space treatment of 
3N Faddeev equations to include in addition to the deuteron also
$^1S_0$ dineutron. Solution of these equations with modified nn
$^1S_0$ CD~Bonn force provided predictions for  cross
section in different nd reactions.

We found that available nd data for the total nd interaction 
 cross section are incompatible with the existence of a
 dineutron with binding energy of absolute value greater than 
$\approx 100$~keV. The data for the total elastic scattering and breakup  
 cross sections do not exclude such a possibilty. 
Also data for the  nd elastic scattering angular distribution 
 cannot decisively exclude such a state. However, in this case the
 precise data taken at forward angles, if available, 
 could provide more constraints  on existence of a dineutron. 

The modifications of the $^1S_0$ nn force component cannot provide
explanation for the drastic discrepancy between theory and data for
the SST geometry of the nd breakup. Allowing for dineutron provides
even smaller SST cross sections, increasing thus that discrepancy. 

The transition from negative to positive nn scattering lengths leads to
drastic modifications of the FSI cross sections. In the outgoing
proton spectra of the uncomplete nd breakup the positive scattering
length  leads to strong
diminishing of the FSI peak at maximal proton energies. 
The carefull Monte Carlo theoretical analysis of existing proton
spectra is required to get answer if these spectra provide a clear signal for
the existence of dineutron. However, complete FSI configurations for which
data have been taken exclude positive values for $a_{nn}$.

\section*{Acknowledgments}
This work was supported by the Polish National Science Center 
 under Grant No. DEC-2011/01/B/ST2/00578. 
 It was also partially supported by 
  the European Community-Research Infrastructure
Integrating Activity
``Exciting Physics Of Strong Interactions'' (acronym WP4 EPOS)
under the Seventh Framework Programme of EU.
 The numerical
calculations have been performed on the
 supercomputer cluster of the JSC, J\"ulich, Germany.

\clearpage
\newpage

\begin{table} 
\caption{The dineutron binding energy $\epsilon_{nn}$,  the nn scattering length
  $a_{nn}$ and the effective range parameter $r_{eff}$ for different 
factors $\lambda$ by which the nn $^1S_0$ component of the CD~Bonn
potential was multiplied.}
\label{tab1}       
\begin{tabular}{|l|c|c|c|}
\hline\noalign{\smallskip}
$~\lambda$  &~ $\epsilon_{nn}$ [MeV]~ &~ $a_{nn}$ [fm]~ &~ $r_{eff}$ [fm]~ \\ 
\noalign{\smallskip}\hline\noalign{\smallskip}
0.9   &   -       &  -8.25  & 3.12    \\
1.0   &   -       &  -18.80   & 2.82    \\
1.19  &   -0.099  &  +21.69   & 2.39   \\
1.21  &   -0.144  &  +18.22  & 2.35    \\ 
1.3   &   -0.441  & +10.95  &  2.20   \\ 
1.4   &   -0.939  &  +7.87   &  2.07  \\ 
\noalign{\smallskip}\hline
\end{tabular} 
\end{table}

\begin{table}
\caption{The theoretical (evaluated at the nucleon
  laboratory  energy $E_{th}$) and experimental (taken at $E_{exp}$)
  nd total cross sections. Theoretical values were obtained with the
  CD~Bonn potential the $^1S_0$ nn component of which was 
multiplied by a   factor $\lambda$.}
\label{tab2}       
\begin{tabular}{|l|c|c|c|c|c|c|c|}
\hline\noalign{\smallskip}
 ~$E_{th} $  & $\sigma_{exp}$ & $E_{exp} $ &
$\sigma_{th}^{\lambda=0.9}$ & $\sigma_{th}^{\lambda=1.0}$ & 
$\sigma_{th}^{\lambda=1.21}$ & $\sigma_{th}^{\lambda=1.3}$ & 
$\sigma_{th}^{\lambda=1.4}$ \\
 ~[MeV]  & [mb] & [MeV]  & [mb] & [mb] & [mb] & [mb] & [mb] \\
\noalign{\smallskip}\hline\noalign{\smallskip}
8.0   & 1207 $\pm$ 13    & 8.0~\cite{Davis} & 1203.4 & 1205.6 & 1258.5
& 1301.4 & 1353.5 \\ 
      & 1213.3 $\pm$ 5.58    & 8.038~\cite{clement} & & & & &  \\
      & 1224 $\pm$ 10    & 8.0~\cite{Schwarz} & & & & &  \\
\hline\noalign{\smallskip}
10.0   & 1055 $\pm$ 10    & 10.0~\cite{Davis} & 1026.4 & 1036.1 &
1089.9 & 1123.7 & 1162.5 \\ 
      & 1051.1 $\pm$ 6.9    & 10.026~\cite{clement} & & & & &  \\
      & 1045.0 $\pm$ 3.4127    & 9.9218~\cite{LANL} & & & & &  \\
\hline\noalign{\smallskip}
13.0   & 867 $\pm$ 12    & 12.995~\cite{Davis} & 837.96 & 851.76 &
900.91 & 926.56 & 954.72 \\ 
\noalign{\smallskip}\hline\noalign{\smallskip}
14.1  & 803 $\pm$ 14    & 14.1~\cite{Poss} & 783.94 & 798.25 & 845.37
& 868.91 &  894.52 \\ 
      & 790 $\pm$ 20    & 14.1~\cite{Cook} & & & & &  \\
      & 809 $\pm$ 6    & 14.1~\cite{Koori72} & & & & &  \\
      & 778 $\pm$ 22    & 14.1~\cite{Shirato} & & & & &  \\ 
      & 806 $\pm$ 6    & 14.1~\cite{Shirato} & & & & &  \\
      & 810 $\pm$ 30    & 14.2~\cite{Meyer} & & & & &  \\
\hline\noalign{\smallskip}
19.0  & 627.96 $\pm$ 12.16    & 18.932~\cite{clement} & 603.47 &
617.55 & 655.92 & 673.20 & 691.76  \\ 
      & 632 $\pm$ 14    & 19.01~\cite{SeagraveHenkel} & & & & &  \\
\hline\noalign{\smallskip}
26.0  & 455 $\pm$ 12    & 26.015~\cite{Davis} & 444.41 & 456.18 &
485.43 & 497.83 & 511.16 \\ 
      & 451.47 $\pm$ 17.72    & 26.082~\cite{clement} & & & & &  \\ 
\hline\noalign{\smallskip}
42.5  & 267.7 $\pm$ 3.9    & 42.5~\cite{Riddle} & 259.24 & 266.35 &
283.32 & 290.27 & 297.88 \\ 
\hline\noalign{\smallskip}
65.0  & 166.5 $\pm$ 2.9    & 63.5~\cite{Riddle} & 157.24 & 160.95 &
170.27 & 173.96 & 178.13 \\ 
      & 161.7 $\pm$ 2.8    & 66.5~\cite{Riddle} & & & & &  \\ 
      & 168.27.0 $\pm$ 0.48333    & 65.039~\cite{LANL} & & & & &  \\
\hline\noalign{\smallskip}
\end{tabular}
\end{table}

\begin{table} 
\caption{The doublet $^2a_{nd}$ and quartet $^4a_{nd}$ nd scattering lengths 
 for different 
factors $\lambda$ by which the nn $^1S_0$ component of the CD~Bonn
potential was multiplied. The calculations have been done with all partial
waves with 2N total angular momenta up to $j_{max}=3$ included.}
\label{tab3}       
\begin{tabular}{|l|c|c|c|}
\hline\noalign{\smallskip}
$~\lambda$  &~ $^2a_{nd}$~[fm]~ &~ $^4a_{nd}$~[fm] \\ 
\noalign{\smallskip}\hline\noalign{\smallskip}
0.9   &   1.51485     &  6.34602  \\
1.0   &   0.93174     &  6.34600  \\
1.21  &  -0.43567     &  6.34596  \\ 
1.3   &  -1.18887     &  6.34593  \\ 
1.4   &  -2.37605     &  6.34589  \\ 
\noalign{\smallskip}\hline
\end{tabular} 
\end{table}

\newpage

\end{document}